\documentclass[aps,pra,twocolumn,showpacs,amsmath,amssymb,preprintnumbers,superscriptaddress,10pt, longbibliography]{revtex4-1}

\usepackage{bm}
\usepackage{scrextend}
\usepackage{graphicx}
\graphicspath{{figures/}}
\usepackage{SIunits}
\usepackage{hyphenat}
\usepackage[bookmarks=false]{hyperref}
\usepackage{color}
\usepackage[capitalise]{cleveref}
\crefname{section}{Sec.}{Secs.}
\Crefname{section}{Section}{Sections}
\usepackage{multirow}
\usepackage{makecell}

\definecolor{pink}{RGB}{255,0,255}

\definecolor{darkcyan}{RGB}{0,174,179}

\begin{document}

\title{Laser-damage attack against optical attenuators in quantum key distribution}

\author{Anqi~Huang}
\email{angelhuang.hn@gmail.com}
\affiliation{Institute for Quantum Information \& State Key Laboratory of High Performance Computing, College of Computer, National University of Defense Technology, Changsha 410073, People's Republic of China}
\affiliation{Institute for Quantum Computing, University of Waterloo, Waterloo, ON, N2L~3G1 Canada}
\affiliation{\mbox{Department of Electrical and Computer Engineering, University of Waterloo, Waterloo, ON, N2L~3G1 Canada}}

\author{Ruoping~Li}
\affiliation{Department of Physics and Astronomy, University of Waterloo, Waterloo, ON, N2L~3G1 Canada}

\author{Vladimir~Egorov}
\affiliation{Faculty of Photonics and Optical Information, ITMO University, 199034 Kadetskaya line 3b, St.~Petersburg, Russia}

\author{Serguei~Tchouragoulov}
\affiliation{QGLex Incorporated, 105 Schneider Road, Suite 111, Ottawa, ON, K2K~1Y3 Canada}

\author{Krtin~Kumar}
\affiliation{Institute for Quantum Computing, University of Waterloo, Waterloo, ON, N2L~3G1 Canada}

\author{Vadim~Makarov}
\affiliation{Russian Quantum Center, Skolkovo, Moscow 121205, Russia}
\affiliation{\mbox{Shanghai Branch, National Laboratory for Physical Sciences at Microscale and CAS Center for Excellence in} \mbox{Quantum Information, University of Science and Technology of China, Shanghai 201315, People's Republic of China}}
\affiliation{NTI Center for Quantum Communications, National University of Science and Technology MISiS, Moscow 119049, Russia}
\affiliation{Department of Physics and Astronomy, University of Waterloo, Waterloo, ON, N2L~3G1 Canada}

\date{\today}

\begin{abstract}
Many quantum key distribution systems employ a laser followed by an optical attenuator to prepare weak coherent states in the source. Their mean photon number must be pre-calibrated to guarantee the security of key distribution. Here we experimentally show that this calibration can be broken with a high-power laser attack. We have tested four fiber-optic attenuator types used in quantum key distribution systems, and found that two of them exhibit a permanent decrease in attenuation after laser damage. This results in higher mean photon numbers in the prepared states and may allow an eavesdropper to compromise the key.
\end{abstract}

\maketitle

\section{Introduction}
\label{sec:intro}

Ideally, quantum key distribution (QKD)~\cite{bennett1984,bennett1992} promises information-theoretic security owing to the solid foundation of quantum mechanics. However, there are often implementation flaws and equipment imperfections, which can be exploited by an eavesdropper Eve to reduce the security of the secret key. This attempt is referred to as ``quantum hacking", drawing the parallel with traditional cybersecurity. The capability of Eve to compromise the security of QKD systems has previously been shown~\cite{lamas-linares2007,nauerth2009,xu2010,lydersen2010a,sun2011,jain2011,tang2013,bugge2014,sajeed2015,huang2016,makarov2016,sajeed2016,huang2018,pang2019,chistiakov2019,gras2019,chaiwongkhot2019}, especially against single-photon detectors~\cite{makarov2006,lydersen2010a,lydersen2010b,wiechers2011,weier2011,lydersen2011b,lydersen2011c,sauge2011,sajeed2015a,sajeed2016,huang2016,chistiakov2019,gras2019,chaiwongkhot2019}, and also was field-demonstrated~\cite{gerhardt2011} with current technology.

To protect a QKD system from detector-side loopholes, measurement-device-independent QKD (MDI QKD) has been proposed~\cite{lo2012} and implemented not only in the laboratory~\cite{rubenok2013, yin2016} but also in the field~\cite{tang2015,tang2016}. MDI QKD removes the security assumption about the measurement station, which can even be an untrusted party~\cite{lo2012}. However, the sources, are still assumed to be in secure laboratories, which might not be true in a practical scenario. Side channels may still exist during quantum-state preparation. Therefore, it is important to further investigate practical vulnerabilities of the source station, to be able to correct and improve its security.

A QKD system often employs a weak coherent laser as a source, with a mean photon number attenuated to single-photon levels using an optical attenuator. Thus, the majority of non-empty pulses contain a single photon, which cannot be split off by Eve and measured separately. There is however a small portion of multi-photon pulses from the inherent Poissonian statistics behind the optical attenuation. The side effect of the multi-photon states can be eliminated by applying decoy-state protocol~\cite{lo2005,ma2005,wang2005a}. However, if the optical attenuation component itself can be altered and its attenuation decreased, either permanently or temporarily, the assumption about the mean photon number may be broken. Eve can then compromise the security of the QKD system~\cite{brassard2000,felix2001,pereira2018,huang2019,zheng2019,zheng2019a}. In particular, even a very small increase of the mean photon number requires a correction to the secret key rate in decoy-state BB84 and MDI QKD protocols, otherwise the key becomes insecure~\cite{huang2019}.

In a source's apparatus, an optical attenuator is usually the last component that optical pulses go through before they are sent to a quantum channel~\cite{takesue2007,inoue2009,yuan2009,liu2013,tang2014a,tang2015,wang2015,fu2015,gleim2016,gleim2017} However, for Eve, the attenuator is the first component she sees, looking at the source's apparatus from the network side. High-power laser damage of other components in QKD systems has been demonstrated before \cite{bugge2014,makarov2016}. We suspect that a high-power laser shining through the output fiber into the source can affect the performance of the attenuator. The changes in its characteristics will then occur before those in other components, because the attenuator dissipates most of the power (assuming a reasonably equal threshold of failure in components). We have selected four types of optical attenuators from four different QKD systems, in each of which the attenuator is positioned to be the last component in the transmitter and closest to the quantum channel. Here we attempt to alter their performance by laser damage. This has been done in an optical setup similar to a live QKD system.

The Article is structured as follows. We first discuss the power handling capability of single-mode fiber in~\cref{sec:high-power}, which gives us an estimate of the maximum power we can apply in the experiment and also in a live system. \Cref{sec:methodology} introduces the experimental setup and methodology. The testing results are presented in~\cref{sec:results}. Briefly, one type of attenuator has survived our testing intact, another exhibited a small temporary attenuation drop, and the remaining two types exhibited a significant and permanent decrease in attenuation. We update the statistical risk prediction for untested QKD systems in \cref{sec:statistics}, discuss ideas for countermeasures in \cref{sec:countermeasures}, and conclude in \cref{sec:conclusion}.

\section{Optical power handling capacity of single-mode fibers}
\label{sec:high-power}

As the first step, we should clarify that how much optical power can be transmitted through a single-mode fiber. Here we set a restriction that Eve can use only the standard single-mode fiber as in a typical QKD system. The amount of optical power that can be sent to the attenuator in the source is limited by the inherent handling capability of the single-mode fiber, as well as the maximum power of the laser used in the attack.

A laser-induced damage threshold (LIDT) of the standard single-mode fiber is theoretically limited by the softening point of silica and its tolerance to thermally induced stress \cite{wood2003}. However, in reality, thermal damage likely happens at the fiber connection points or the interface between the fiber core and the cladding~\cite{Yanagi2002,wood2003}. A fiber fuse phenomenon can be triggered by high temperatures at a fiber end facet~\cite{kashyap1988,kashyap2013}. This can be reproduced by contacting the fiber end against an absorptive material, such as metal, or by using a flame ($\sim 2700~\celsius$)~\cite{davis1997}. It has been experimentally demonstrated that a $2$--$5~\watt$ continuous-wave (c.w.)\ laser can initiate the fiber fuse~\cite{kashyap1988,davis1997}. However, in our testing of a $20$-$\meter$ long single-mode fiber ending with no termination (a $90\degree$ cleave), when no deliberate method was applied attempting to trigger the fiber fuse, it was able to tolerate $9~\watt$ c.w.\ laser, the specified maximum power of our laser source.

Power loss in the fiber limits the power that can be delivered to the target. The major threat comes from backward scatterings in the optical fiber, especially the backward stimulated Raman scattering~(SRS) and stimulated Brillouin scattering~(SBS)~\cite{smith1972}. Generally, during light transmission, a fraction of incident light can be transferred from one optical field to another field with frequency shift, due to molecular vibrations of the transmission medium. The frequency-shifted light is called Stokes wave. The intensity of Stokes wave may rapidly increase over distance, which causes further SRS and SBS. This scattered light can travel backward to the laser source and may destroy it. To keep the high-power laser source safe, the SRS and SBS thresholds need to be confirmed. The threshold is defined as the incident pump power $P_{\text{th}}$ at which the backward Stokes power $P_{\text{s}}$ becomes equal to the power at the fiber output~\cite{smith1972}
\begin{equation}\label{eq:def}
 P_{\text{th}}e^{-\alpha L} = P_{\text{s}}(L), 
\end{equation}
where $\alpha$ is the fiber loss (typically $0.05~\kilo\metre^{-1}$ at $1550~\nano\meter$), and $L$ is the transmission distance in $\kilo\metre$. \Cref{eq:def} indicates that the threshold is dependent on the fiber length.
 
\begin{figure}
	\includegraphics{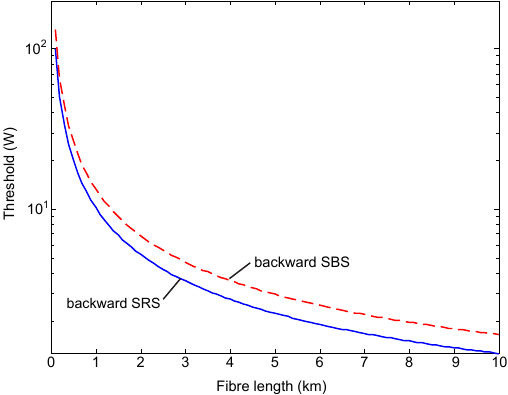}
	\caption{Simulated backward SRS and SBS thresholds. The latter assumes a wideband source; see text for details.}
	\label{fig:threshold}
\end{figure}

For the backward SRS, its threshold is given by~\cite{smith1972}
\begin{equation}\label{eq:SRS}
 P_{\text{th}}^{\text{SRS}} = \frac{20A_{\text{eff}}}{g_{\text{R}} L_{\text{eff}}}.
\end{equation}
Here $A_{\text{eff}}$ is the effective core area (for the standard single-mode fiber with the core diameter of $8~\micro\meter$, $A_{\text{eff}} = 50~\micro\meter\squared$); $g_{\text{R}}$ is the Raman-gain coefficient, which is $6.67\times10^{-14}~\meter/\watt$ at $1550~\nano\meter$~\cite{agrawal2007}; $L_{\text{eff}}$ is the effective interaction length defined as~\cite{agrawal2007}
\begin{equation}\label{eq:length}
 L_{\text{eff}} = \frac{1-e^{-\alpha L}}{\alpha}.
\end{equation}
Thus, the threshold value is dependent on the transmission distance. The simulation result of $P_{\text{th}}^{\text{SRS}}$ versus transmission distance is given in~\cref{fig:threshold}, which shows that the SRS threshold drops dramatically when the fibre length extends. However, more than $10~\watt$ optical power is allowed for transmission distance shorter than $1~\kilo\meter$.

On the other hand, the backward SBS plays a key role in limiting the transmission power. SBS can occur at much lower incident power level than that for SRS. The input power threshold is quantified by~\cite{smith1972}
\begin{equation}\label{eq:SBS}
 P_{\text{th}}^{\text{SBS}} = \frac{21A_{\text{eff}}}{g_{\text{B}} L_{\text{eff}}},
\end{equation}
where the Brillouin-gain coefficient $g_{\text{B}} = 5\times10^{-11}~\meter/\watt$~\cite{agrawal2007}. This only allows $2.1~\watt$ as maximum input power at $L=10~\metre$. Fortunately, the SBS threshold can increase considerably if the spectral width of the pump laser $\Delta\nu_p$ is much larger than the full width at half maximum (FWHM) of the Brillouin-gain spectrum in SBS $\Delta\nu_B$~\cite{agrawal2007}. Then the SBS threshold in~\cref{eq:SBS} increases by a factor of $1+\Delta\nu_p/\Delta\nu_B$. In the experiment, we employ a laser diode with $\Delta\nu_p \approx 10~\giga\hertz$; $\Delta\nu_B$ in the single-mode fiber at $1550~\nano\meter$ is $16~\mega\hertz$. The SBS threshold in this case is also shown in~\cref{fig:threshold}. In this case it is slightly higher that the SRS threshold, i.e.,\ the latter remains the limiting factor.

Overall, at least $10~\watt$ c.w.\ laser power can safely be transmitted through $1~\kilo\meter$ single-mode fiber. In most network installations we expect Eve to be able to connect to the quantum channel within this distance of the source. Shorter distance translates to larger fiber handling capability of c.w.\ power, which translates to larger power that Eve can apply to the attenuator in the source. In our experiment, we use an amplified laser source providing up to $9~\watt$ c.w.\ power and transmit it through a $20~\meter$ long fiber. The feasibility of longer transmission is theoretically verified by the models above.

\section{Testing method}
\label{sec:methodology}

\subsection{Experimental setup}
\label{sec:setup}

\begin{figure}
	\includegraphics{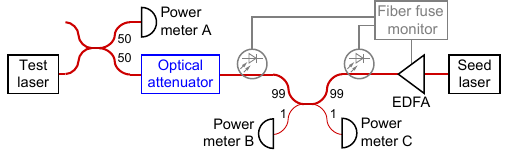}
	\caption{Simplified diagram of experimental setup, with the optical attenuator as a replaceable device under test. The output of erbium-ytterbium doped fiber amplifier (EDFA) is fusion-spliced to the 99\% arm of the fiber beamsplitter. All fibers are standard single-mode.}
	\label{fig:setup}
\end{figure}

The test of the optical attenuators has been conducted using the setup shown in~\cref{fig:setup}. The experimental scheme mimics a hacking scenario for a running QKD system. The test laser is a fiber-pigtailed 1550~nm laser diode~(Gooch \& Housego AA1406), acting as the laser in the source. This laser provides $5~\milli\watt$ c.w.\ light to measure the attenuation of the optical attenuator under test. The input of this attenuator is connected to the test laser through a $50\!:\!50$ fiber beamsplitter~(BS). Power meter~A (Joinwit JW3208) monitors the power of the test laser, and power meter~C (Thorlabs PM200 with S154C sensor) serves to check the attenuation of the optical attenuator before and after optical damage. An erbium-ytterbium doped fiber amplifier (EDFA) provides up to $9~\watt$ c.w.\ power, which is applied to the optical attenuator through a $99\!:\!1$~BS~(Thorlabs 10202A-99-FC). A fiber-pigtailed laser diode emitting in a broad $1550.06$--$1550.14~\nano\meter$ band (QPhotonics QFBGLD-1550-100) set at $20~\milli\watt$ c.w.\ power is used as a seed source for the EDFA. Power meter B (Thorlabs PM200 with S146C sensor) measures the optical power at the EDFA's output. This scheme mimics a live scenario in which Eve injects light into QKD source station via the quantum channel.

The attenuation of the optical attenuator is determined by comparing the readings from the power meters A and C, taking into account the additional $20~\deci\bel$ attenuation of the $99\!:\!1$~BS. This measurement is first performed for each attenuator with the optical amplifier and seed laser turned off, so that the initial attenuation is calibrated before any attempted tampering. The same measurement is repeated again after each laser damage test. Any attenuation change after activating the EDFA can then be attributed to optical damage within the setup. We have verified separately that during and after our testing up to the maximum $9~\watt$ power, the beamsplitters do not significantly change their splitting ratio. Thus any measured attenuation change is localised in the optical attenuator. In case bulkhead fiber-optic connectors get burnt (that some of the tested attenuator types use), we treat it as an outcome of Eve's attack that would cause a denial of service in the actual QKD system~\cite{makarov2016}.

Our high-power light source in this experiment is the custom-made $1550~\nano\meter$ erbium-ytterbium doped fiber amplifier, manufactured by QGLex. It is designed using core pumping at the first stage and double-cladding pumping at the second stage. This allows for high gain and can accommodate the input seed power as low as $0.4~\milli\watt$, which is amplified to a maximum of about $9~\watt$ ($39.5~\deci\bel\milli$) through a standard single-mode fiber (SMF-28) with a high slope efficiency of 28\%. This amplifier has additional suppression of amplified spontaneous emission (ASE) at $1.0~\micro\metre$. A significant presence of ASE could lead to spurious lasing at $1.0~\micro\metre$ and limitation of the energy transfer process from the ytterbium ions to the erbium, which would limit the output power level and efficiency. The EDFA exhibits single mode behaviour and a high slope efficiency. Before our experiment, the output power of the EDFA has been calibrated using power meter~B (\cref{fig:setup}). This calibration provides an accurate relation between the software setpoint for the EDFA and its actual output power. 

\begin{table*}
\vspace{-0.8em} 
\caption{Summary of test results for all attenuator samples. For the last attenuator type the number of samples means the number of individually tested points across the variable-density disk of a single device sample. For variable attenuators, the average change in attenuation after damage $\Delta$ is measured at their attenuation setpoint used during the laser damage. ``Safe'' power does not cause observable changes in attenuation. Attack threshold is the laser power at which the attenuator starts to exhibit more than $1~\deci\bel$ drop in attenuation. Failure threshold denotes the power at which the attenuator begins to fail catastrophically, increasing its attenuation by more than $3~\deci\bel$. \vspace{0.8\baselineskip}}
\footnotesize
\begin{tabular}{l@{\enskip}l|c@{\enskip}c@{\enskip}ccccc}
 \multirow{2}{*}{Type}
 & \multirow{2}{*}{\makecell{Manufacturer\\ and part number}}
 & \multicolumn{3}{c}{Number of samples}
 & \multirow{2}{*}{\makecell{Maximum ``safe''\\ power (dBm)}}
 & \multirow{2}{*}{\makecell{Average success\\ $\Delta$ (dB)}}
 & \multirow{2}{*}{\makecell{Average attack\\ threshold (dBm)}}  
 & \multirow{2}{*}{\makecell{Average failure\\ threshold (dBm)}} \\
 & & Total & Success & Failure \\
 \hline
 Manual VOA & \makecell[l]{OZ~Optics BB-700-11-1550-\\8/125-P-60-3A3A-1-1-LL} & 2 & 0 & 0 &\textgreater39.5\footnote{\label{ft:maxpower}$39.5~\deci\bel\milli$ is the maximum power tested, limited by our EDFA.} & -- & -- & -- \\ 
 Fixed & unspecified\footnote{\label{ft:withheld}Information withheld at the request of QKD manufacturer.} & 12 & 4 & 6 & 32.8& $-1.37$ & 34.0 & 37.2 \\
 MEMS VOA & unspecified\footref{ft:withheld} & 13 & 8 & 4 &34.5& $-5.34$ & 36.2 & 36.6 \\
 VDMC VOA & FOD 5418 & (25) & (18) & 0 &32.9& $-9.59$ & 34.5 & 36.5 \\
\end{tabular}
\label{tab:all}
\end{table*}

In case a fiber fuse \cite{kashyap1988,kashyap2013} occurs in the device under test while applying high power during the experiment, an automatic monitoring and shutdown system protects the rest of the setup. We have implemented it with two Si photodiode sensors placed laterally along the fiber jacket. The sensors detect thermal visible light emitted by hot plasma in the fiber fuse as it propagates past them. One sensor is placed at the attenuator output, and another at the output of the EDFA~(\cref{fig:setup}). If either of them detects light, a monitoring circuit shuts down the pump in the EDFA. This shuts off the emission and stops the fiber fuse at the sensor. This circuit has activated several times during our tests, preventing extensive damage to the equipment.

\subsection{Test procedure}
\label{sec:procedure}

We can define a successfully ``hacked" sample as one having a drop in attenuation after optical damage. For variable attenuators, this drop needs to be observed within their range of attenuation setpoints used in a QKD system. To quantify the result, we set a threshold of $1~\deci\bel$ drop in attenuation (which we can reliably measure above experimental errors) after damage, beyond which the attack is deemed to be successful. This drop approximately corresponds to $26\%$ increase in the mean photon number within the quantum channel. We also arbitrarily define a critical attack failure as the situation when the attenuator exhibits an attenuation increase larger than $3~\deci\bel$, corresponding to a drop in the mean photon number of about $50\%$.

For each optical attenuator sample, our testing procedure is the following. The attenuation of the sample under test is set to a value that is typically used in each specific QKD system that employs this type of attenuator. The test laser is always on. The EDFA applies high power starting at $316~\milli\watt$ ($25~\deci\bel\milli$) for at least $10~\second$. Afterward, the EDFA is turned off and the attenuation is measured. If no change in the attenuation has occurred, the high power is increased by $0.5$--$1~\deci\bel\milli$ and the steps above are repeated. Once a change in attenuation ($< -1~\deci\bel$ or $> 3~\deci\bel$) is detected, the testing is stopped. If the maximum EDFA power of $9~\watt$ is applied with no change in attenuation, the testing is also stopped. We have observed that the attenuators heat up during the high-power exposure and take time to cool down, during which temporary changes in attenuation have been recorded for some samples. A permanent change in attenuation may then remain after it has fully cooled down to the room temperature.

If the decrease of attenuation is observed in one sample, the test is repeated on additional samples of the device to demonstrate the effect more than once.

\section{Experimental results}
\label{sec:results}

A total of four types of optical attenuators have been tested. Of these, one type appears to be minimally affected by optical damage up to $9~\watt$ ($39.5~\deci\bel\milli$); another type consistently shows a temporary decrease in attenuation. The two remaining types exhibit permanent changes in attenuation after being subjected to the high-power laser. A summary of the laser damage results is presented in \cref{tab:all}.

\subsection{Manual variable attenuator}

The first attenuator is a manually adjustable type (OZ~Optics BB-700-11-1550-8/125-P-60-3A3A-1-1-LL) with a range of $1.5$ to $80~\deci\bel$. It consists of a miniature fiber bench with two collimating lenses which expand the beam to $\sim 1~\milli\meter$ diameter and couple it back into the fiber. The collimated beam is partially obstructed by the opaque tip of a metal screw. Rotating the screw adjusts how much it is inserted into the beam, thus adjusting the amount of attenuation.

We have tested this attenuator set at $31~\deci\bel$, corresponding to an almost complete blocking of the beam with the screw. The attenuator's polarisation-maintaining fiber pigtail has been spliced to the single-mode fiber at the high-power side, with about $1~\deci\bel$ splice loss, which mimics the way it is connected in the QKD system that employs it. This PM-to-SM connection is not expected to significantly affect the test, because most of the high power is still delivered to the attenuator. The latter is not polarisation-selective.

We have not observed any change in the attenuation even at the highest available laser power of $9~\watt$ ($39.5~\deci\bel\milli$) applied continuously for $20~\minute$. The attenuator case reached an equilibrium temperature of $234~\celsius$ at this power, as measured using a thermal imaging camera. This has led to discoloration of the black anodised coating on the aluminum case, and deformation of plastic strain relief sleeves. 

A closer visual inspection of the attenuator has been performed to determine the damage mechanism. Upon disassembly, the optical blocking material at the adjustable screw tip reveals a concave dent facing the input fiber where the laser power has been delivered. This suggests that the blocking material is being damaged and removed by the laser, likely through vaporisation or ablation of the material under high temperature. Consequently, under a higher c.w.\ optical power or ablation from a pulsed laser, there might be further damage to the screw with the possibility of a complete perforation. Then a permanent decrease in attenuation might occur. In short, our testing of this attenuator type has been inconclusive, being limited by our experimental capabilities, but a higher power may decrease its attenuation.

\subsection{Fixed attenuator}

The second type is a fixed attenuator of nominal attenuation $25~\deci\bel$. The attenuator is a short cylindrical module having a male FC/PC connector at one side and female one at another side. Its physical disassembly shows a solid axial cylinder of dark ceramic material roughly 5~mm long placed inline between the input and output fibers, which absorbs incident light. This absorber is placed in a white ceramic ferrule, which is in turn placed in a series of concentric metal sheaths beneath the outer casing.

\begin{figure}
	\includegraphics{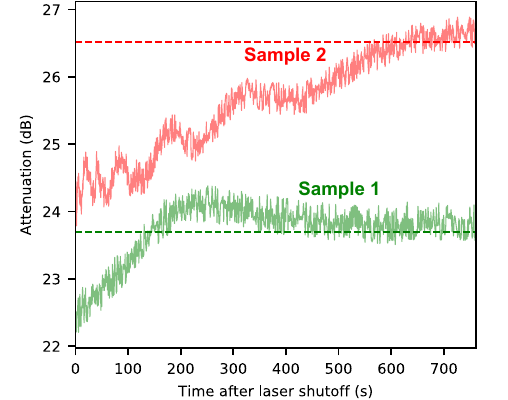}
	\caption{Behaviour of the fixed attenuator immediately after being subjected to a high-power laser at $4~\watt$~(36~dBm) for $5~\minute$. A temporary decrease in attenuation is observed until the attenuator cools down in $\sim 10~\minute$. The horizontal lines indicate initially measured attenuation before the application of high power.}
	\label{fig:fixed-ATT}
\end{figure}

Illumination power up to $1.91~\watt$ ($32.8~\deci\bel\milli$) does not affect our fixed attenuator samples. However, immediately after the application of approximately $4~\watt$ ($36~\deci\bel\milli$) power, they exhibit a temporary decrease in attenuation. The maximum decrease of approximately $2~\deci\bel$ occurs in one of the two samples at laser shutoff, as shown in~\cref{fig:fixed-ATT}. The attenuation then reverts back to the initial state within minutes. It is however still possible for Eve to exploit this temporary decrease in attenuation, leaking parts of the secret key~\cite{brassard2000,felix2001,sajeed2015,pereira2018,huang2019}. This vulnerability window then exists only for a limited time after the high-power exposure. At a higher power of approximately $6.3~\watt$ ($38~\deci\bel\milli$), critical damage occurs to the absorptive element of the device, with the attenuation permanently increased by more than $20~\deci\bel$. 

To explain the possible damage mechanisms, we hypothesise that either the absorbing material or the attenuator housing can be made to expand and contract under thermal stress, which would explain the temporary drop in attenuation, as small amount of optical power can leak to the output between the casing and the absorbing material. For samples suffering from critical damage, disassembly shows that the dark ceramic absorbing material has a burned appearance, with irregular dark patches surrounding it. The critical damage that produces a large increase in attenuation ($20~\deci\bel$) likely corresponds to when this ceramic material is darkened under the high temperature generated by the illumination.

\subsection{MEMS-based variable attenuator}

\begin{figure}
	\includegraphics{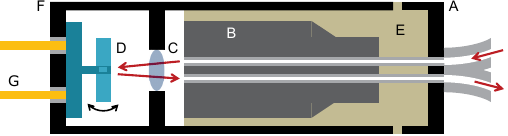}
	\caption[Simplified cross-section of MEMS VOA]{Simplified schematic of MEMS variable optical attenuator, with parts labeled (not to scale). A,~metal cap holding the input and output single-mode fibers; B,~glass sheath; C,~collimating lens; D,~voltage adjustable MEMS mirror on torsion mount; E,~adhesive filler; F, metal body; G, electrical leads.}
	\label{fig:VOA}
\end{figure}

The third type is a variable optical attenuator (VOA) based on a micro electro-mechanical system (MEMS) element. The MEMS~VOA is voltage-adjustable from a maximum of 31--$34~\deci\bel$ to a minimum of approximately $1~\deci\bel$. Its physical disassembly shows two parallel input and output fibers facing a reflector-lens assembly (\cref{fig:VOA}). The voltage controls the tilt of the reflector mirror via electrostatic force, and thus changes the amount of coupling between the input and output fiber. During the tests, the attenuation of the samples is set within a range that might be used in the QKD system that employs this attenuator.

\subsubsection{Preliminary tests on individual attenuators}

\begin{figure}
	\includegraphics{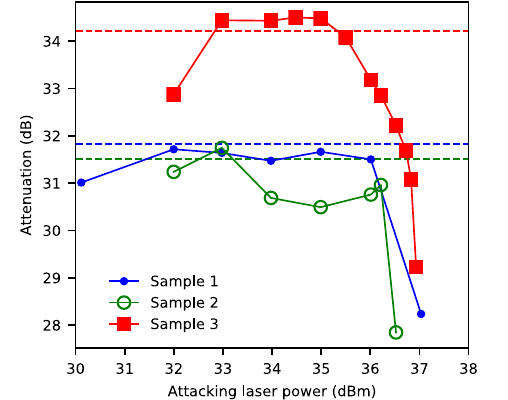}
	\caption{Selected samples of MEMS VOA with permanent decrease in attenuation after laser damage. The attenuation decreases abruptly near $4~\watt$ ($36~\deci\bel\milli$). Horizontal lines indicate the initial measured attenuation before the application of high power. The leftmost point shown is the first application of high power to each sample.}
	\label{fig:MEMS-VOA}
\end{figure}

We have tested 8 samples of MEMS VOAs in an unpowered state, in which they have their maximum attenuation value. Out of these tested, 3 samples exhibit an average change of $-3~\deci\bel$ after optical damage, as shown in~\cref{fig:MEMS-VOA}, which is confirmed to be permanent with subsequent measurements after several hours and days. The permanent decrease in attenuation occurs near laser power of $4~\watt$ ($36~\deci\bel\milli$). Near this damage threshold, the attenuation fluctuates when measured after turning off the EDFA, steadily decreases, and stabilizes after a few minutes. This is likely due to thermal effects. If the failure threshold is exceeded, the optical attenuator sustains catastrophic damage and its attenuation is permanently increased to $>70~\deci\bel$.

\subsubsection{Further tests on assembled attenuator boards}

\begin{figure}
	\includegraphics{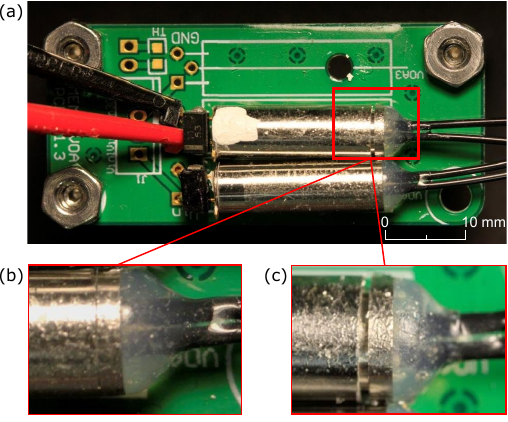}
	\caption[MEMS VOA overview]{MEMS VOAs mounted on (a)~PCB. (b)~Initially, the end cap tightly inserted into the body. (c)~Catastrophic structural damage in MEMS VOA sample~3 at $5.6~\watt$~(37.5~dBm). The cap is displaced.}
	\label{fig:cap}
\end{figure}

Following the initial confirmation that MEMS VOAs are especially vulnerable to the laser damage attack and exhibit a permanent attenuation drop after optical damage, subsequent experiments have been done using complete printed circuit board (PCB) mounted attenuator assemblies received from the industry, shown in~\cref{fig:cap}. The attenuators are from two different manufacturers but are nearly identical in their construction, here labeled subtype A or B, and are attached using a soft silicone glue to the PCB, providing a thermally accurate representation of their use in the QKD system.

\begin{figure}
	\includegraphics{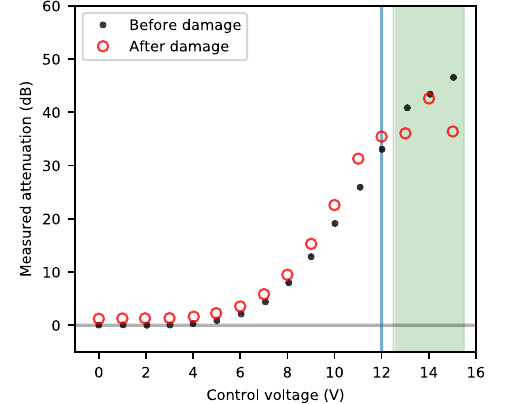}
	\caption{Typical voltage-attenuation curve of a successfully compromised sample~1 of MEMS VOA. Blue vertical line denotes the voltage setting at which the laser damage is done. The shaded area denotes the range where permanent attenuation drop is observed.}
	\label{fig:MEMS-VOA-assembly}
\end{figure}

\begin{figure}
	\includegraphics{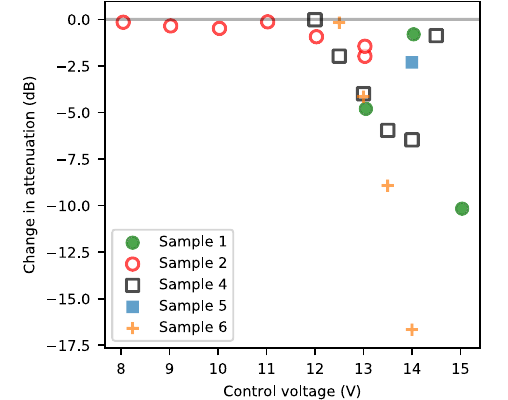}
	\caption{Attenuation change after optical damage for successfully compromised MEMS VOA samples. For clarity, only control voltage points for which the attenuation has decreased are plotted. I.e., the plotted data points correspond to points within the shaded area in \cref{fig:MEMS-VOA-assembly}.}
	\label{fig:MEMS-VOA-sample}
\end{figure}

\begin{table*}
\vspace{-0.8em} 
\caption{Optical damage results for complete MEMS VOA assemblies. Testing voltage is the attenuator control voltage at which high power is applied. $\Delta$ is the change in attenuation observed at the testing voltage.\vspace{0.8\baselineskip}}
\footnotesize
\renewcommand{\tabcolsep}{1mm}
\begin{tabular}{cc|cccccc} 
 Sample
 & Subtype
 & \makecell{Testing\\ voltage (V)} 
 & \makecell{Attenuation\\ before (dB)}  
 & \makecell{Attenuation\\ after (dB)}
 & \makecell{$\Delta$\\ (dB)}
 & \makecell{Attack threshold\\ (dBm)}\\[1.5ex]
 \hline
 1 & A & 12.0 & 33.05 & 35.44 & $+2.39$ & 36.5\\ 
 2 & A & 12.0 & 33.88 & 32.95 & $-0.93$ & 37.0\\
 3 & A & 11.5 & 32.81 & 64.28 & $+31.47$ & 37.5 (failure)\\
 4 & B & 14.0 & 38.79 & 32.32 & $-6.47$ & 35.5\\
 5 & B & 14.5 & $\approx68$ \footnote{Measurement near the minimum power range of the power meter.} & 58.82 & $\approx-9.2$ & 36.0\\
 6 & B & 13.5 & 31.21 & 22.29 & $-8.92$ & 34.5\\
\end{tabular}
\label{tab:MEMS-VOA-boards}
\end{table*}

\begin{figure*}
	\includegraphics{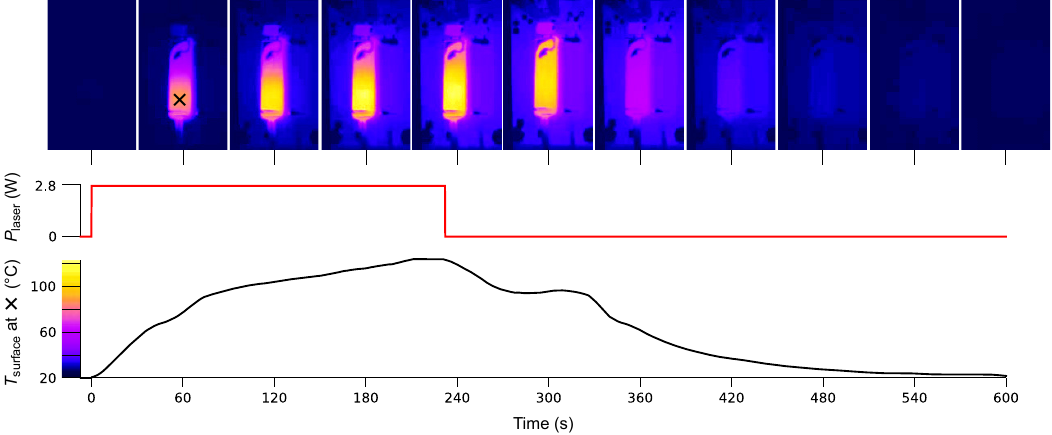}
	\caption{Temperature profile of VOA sample 6 at the attack threshold, taken at a single point (marked \bm{$\times$}) near the cap of the VOA. The high power laser was set to $2.8~\watt$~($34.5~\deci\bel\milli$) and turned on from time 0 through 232~\second.}
	\label{fig:VOA-thermal}
\end{figure*}

Our results are summarized in \cref{tab:MEMS-VOA-boards}. At the testing voltage chosen, corresponding to an attenuation of approximately $30~\deci\bel$, 4 out of the 6 attenuators tested have exhibited a permanent drop in attenuation. Furthermore, for 5 out of the 6 attenuators, there exists an attenuation range with a decrease in attenuation post-damage, as shown by the shaded area in~\cref{fig:MEMS-VOA-assembly}. The response curve of the attenuation within the shaded area is shown for all successfully damaged attenuators in \cref{fig:MEMS-VOA-sample}. We can deem the optical damage attack to be successful within this voltage range. The range at which there is a clear drop in attenuation is almost always at the higher attenuation settings, and~\cref{fig:MEMS-VOA-assembly} shows a typical behavior for the successfully attacked sample. 

Attenuator sample 3 has exhibited a near-total failure, where the attenuation after the optical damage is dramatically increased over its normal value. Effectively, this is similar to a component becoming an open-circuit in electronics, corresponding to the optical component blocking light. This sample represents a case of critical failure, which is an undesirable outcome for Eve resulting in denial of service in QKD.

Attenuator sample 5 is peculiar. When we initially measured its attenuation-voltage curve (before applying high optical power), the attenuation value has become latched at $14.5~\volt$. A subsequent voltage change down to $0~\volt$ did not change this measured attenuation. The $14.5~\volt$ voltage, however, does appear to be in the working range for the other attenuators from manufacturer~B. Since the applied voltage is close to the maximum voltage specified, it is likely that the latching observed at this voltage is from some inherent variability in the working voltage range between components, or a manufacturing defect. However, despite this unexpected malfunction, a permanent decrease in attenuation after laser damage is still observed in this sample.

\subsubsection{Possible damage mechanisms}
\label{sec:reason}

During the application of high power near the damage threshold, the cap holding the input and output fibers bulges outwards (\cref{fig:cap}), possibly pulling the fiber inside the VOA out of alignment with the collimating lens. In a catastrophic damage scenario at approximately $5.6~\watt$~($37.5~\deci\bel\milli$), the cap detaches itself from the attenuator casing and a puff of smoke emits from the opening. ~\Cref{fig:VOA-thermal} shows the thermal profile of the attenuator under high-power laser recorded using an uncooled microbolometer-based thermal camera (FLIR E60). As the images show, the highest temperature occurs near the front end of the VOA casing where the input and output fibers are inserted. Since the process of coupling a beam of light into a single-mode fiber is highly dependent on relative positions of the involved optical elements~\cite{martin1979}, we hypothesize that structural deformation under high temperature is a likely cause for the observed change in attenuation. \\

Another possible cause is that for typical MEMS materials used (Si, SiN, SiC, etc.)~\cite{kotzar2002}, the operating temperature can induce lattice strain and change its spring constant \cite{sharpe2005}. The ductility of polycrystalline Si is reported to increase at temperatures near $500~\celsius$ \cite{sharpe2005}. Since the MEMS micromirror used in the VOA is fixed using a torsion mount, the amount of deflection induced by a given voltage may change with temperature. The voltage-attenuation curve might be different once the VOA heats up and exceeds its proper operating temperature range, which may either result in a drop in attenuation or an increase, depending on the exact material behavior under high temperature. Our observations using the thermal camera show that the outer casing of the VOA reached $120~\celsius$ at the attack threshold power. The internal temperature is likely to be much higher. However, from the observations and physical disassembly, it appears that the area near the fiber end of the VOA is visibly more affected by thermal damage, and not the mirror facet, which attributes some doubt to this mechanism of damage.\\

\subsection{Variable density metal coating variable attenuator}

\begin{figure}
	\includegraphics{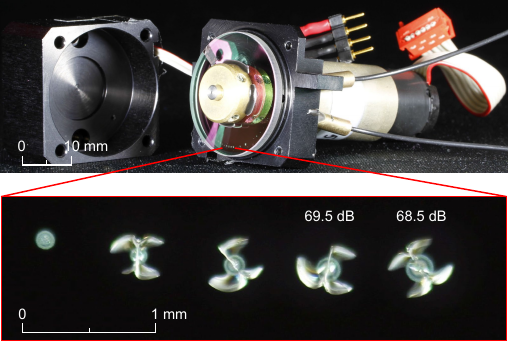}
	\caption{Variable-density metal coating variable attenuator opened with glass disk active element visible. Small dots around the disk edge correspond to irradiated areas. Magnified view of five damaged spots is given in the inset. The two rightmost spots are those shown in \cref{fig:VDMC-1}.}
	\label{fig:VDMC-VOA}
\end{figure}

The fourth type is a programmable VOA (FOD 5418) whose active element is a glass disk covered with variable density metal coating (VDMC), see~\cref{fig:VDMC-VOA}. Collimated light from the input fiber passes through VDMC, the glass, gets reflected from a dielectric mirror deposited at the back surface of the disk, then passes the glass and VDMC again. It then enters another collimator coupling it to the output fiber. An externally controlled motor rotates the glass disk to expose different coating density to the beam, adjusting attenuation in $0$--$80~\deci\bel$ range.

This VDMC~VOA was used in a QKD system developed at ITMO University. The latter was based on a subcarrier-wave architecture with phase protocol and typical mean photon number around $0.2$ \cite{gaidash2019,miroshnichenko2018}. At the time of the experiments described here, the attenuator was the last component before Alice's output \cite{gleim2016,gleim2017}, thus potentially allowing Eve to attack it.

\subsubsection{Testing results}

We have tested one VDMC VOA with a total of 25 measurements over different points of the active element corresponding to its different attenuation value settings, as summarized in \cref{tab:all}. During 7 of these measurements done at laser power $\leq 1.95~\watt$ ($\leq 32.9~\deci\bel\milli$), neither permanent nor temporary change in attenuation has been observed. All the remaining sample points have been successfully ``hacked'', demonstrating a permanent decrease in attenuation with mean change of $-9.59~\deci\bel$. None of the points have been critically damaged when we applied up to $6.8~\watt$ ($38.3~\deci\bel\milli$) over up to $15~\minute$ duration. This was the maximum power available in our test setup at the time of this last experiment, because the EDFA has aged. 

The successful attack threshold for VDMC VOA sample points depends on exposure time. In order to avoid destruction of its active element (due to heat-induced cracking of the glass disk) in our tests, we chose reasonably low exposure time of $10~\second$. After every $10~\second$ of exposure we switched off the EDFA for another $10~\second$ in order to cool down the sample and then resumed our testing at the initial power level. We could therefore observe the added effects of high-power exposure on the metal coating with a minimal risk of sample destruction. In this subsection and in \cref{tab:VDMC-VOA}, the total heat duration is calculated as the sum of all $10~\second$ intervals of high-power exposure, disregarding the cooling time. Our goal was to define optimal conditions in terms of repeatability and minimal total exposure (i.e.,\ both irradiation time and optical power). The latter would potentially decrease the probability of catastrophic sample damage or uncovering Eve's operation. We have found that a minimum power level that leads to the successful attack is $2.0~\watt$~($33.0~\deci\bel\milli$) for $200~\second$ exposure time. At $2.2~\watt$~($33.4~\deci\bel\milli$) the effect is achieved at shorter times varying from 30 to $50~\second$. Finally, exposure for only $10~\second$ at $2.8~\watt$~($34.5~\deci\bel\milli$) has led to a steadily repeatable result over 9~sample points, therefore this power value may be considered a consistent attack threshold. 

\begin{figure}
	\includegraphics{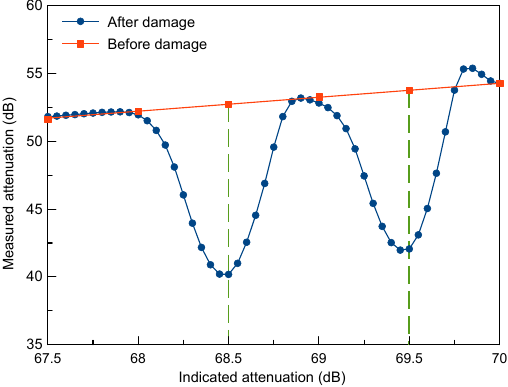}
	\caption{Typical attenuation curve of successfully compromised VDMC VOA. Green vertical lines denote the attenuation settings at which optical damage was done. The high power laser applied $2.8~\watt$~($34.5~\deci\bel\milli$) for $10~\second$.}
	\label{fig:VDMC-1}
\end{figure}

\begin{figure}
	\includegraphics{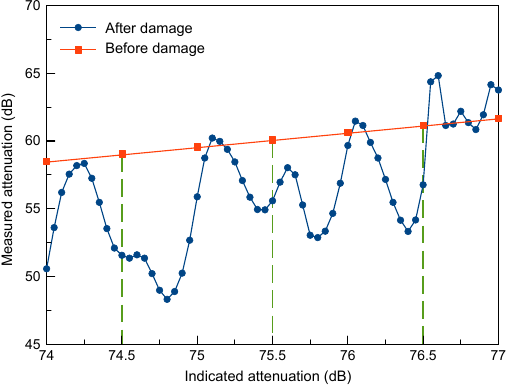}
	\caption{Attenuation curve of VDMC VOA compromised in suboptimal conditions. Green vertical lines denote the attenuation settings at which optical damage was done. High-power exposure parameters are given in the text.}
	\label{fig:VDMC-2}
\end{figure}

Typical attenuation curves before and after successful attacks are shown in \cref{fig:VDMC-1}. ``Indicated attenuation" represents the programmed VOA setting~\footnote{Prior to the experiment, we performed a manual calibration of the VOA measuring its actual attenuation versus the programmed setting over its typical operating region, $20$ to $80~\deci\bel$, with $0.5~\deci\bel$ step. We found that the attenuation curve of our device sample was initially shifted (\cref{fig:VDMC-1}), probably due to a mechanical rotational misalignment of the active element, while it was otherwise a stable and fully functioning device. As the reader may understand, we have chosen this out-of-calibration device sample for our destructive tests, because it had limited value for other purposes. Our conclusions are not affected by the device being out of calibration.}. An area of lowered attenuation has appeared around each affected sample point as the result of laser damage. Optical loss gradually increases around the point and returns to normal after $0.5~\deci\bel$ shift in either direction. \Cref{fig:VDMC-2} illustrates consequences of choosing suboptimal laser damage conditions. When the optical power goes below $2.8~\watt$ (e.g.,\ $74.5~\deci\bel$ point, 2.5~W for $10~\second$), the curve structure becomes asymmetric and attenuation is not minimized at the point of damage. This effect is countered by increasing the exposure time. However, the damage outcome is more sensitive to power than time, as can be seen at the second point in \cref{fig:VDMC-2} ($75.5~\deci\bel$ point, 2.2~W for $50~\second$), where the asymmetry becomes even more prominent. Finally, lowered power and increased exposure time ($76.5~\deci\bel$ point, 2.2~W for $30~\second$) sometimes results in increased loss value around the damaged area and/or a slight shift of the attenuation minimum from the point of damage. We attribute these effects to structural damage of the sample such as the glass surface cracking or the metal coating detaching owing to continuous local overheating.

As explained earlier, the light beam passes through the VDMC twice. We have verified that damaged areas from different sample points never overlapped, ensuring no interference between them in this experiment. To do so, before taking each measurement we checked that attenuator losses coincided with their initial level defined during calibration. Also, the damage observed in most samples was only at the first passage point.

\Cref{tab:VDMC-VOA} summarizes laser damage conditions and attenuation decrease over 10 selected sample points damaged under different conditions. A local minimum value is given for each sample, even when the measurement point did not perfectly coincide with the point of damage (see \cref{fig:VDMC-2}). During a realistic attack Eve cannot control the VOA setting in the source to access this local minimum. Nevertheless, since our experiments indicate that in suboptimal irradiation conditions the acquired attenuation value randomly fluctuates around the damage point, it is reasonable to consider the worst-case scenario when the minimum randomly aligns with the VOA setting. Maximum attenuation decrease in optimal conditions ($2.8~\watt$, $10~\second$) was $-12.58~\deci\bel$. Over all 18 sample points, $-9.54~\deci\bel$ mean change was observed. Minimum change in attenuation ($-2.29~\deci\bel$) was observed in the area of very low initial losses ($5.29~\deci\bel$).  

\begin{table}
\vspace{-0.8em} 
\caption[Optical damage results for VDMC VOA]{Optical damage results for several VDMC VOA sample points.\vspace{0.8\baselineskip}}
\footnotesize
\renewcommand{\tabcolsep}{1mm}
\begin{tabular}{c|ccccc} 
 \makecell{Sample\\ point}
 & \makecell{Attenuation\\ before (dB)} 
 & \makecell{Attenuation\\ after (dB)} 
 & \makecell{$\Delta$\\ (dB)}
 & \makecell{Power\\ (W)}
 & \makecell{Total heat\\ duration (s)}\\[1.5ex]
 \hline
 1 & 5.29 & 3.0 & $-2.29$ & 2.81 & 10\\ 
 2 & 9.86 & 4.21 & $-5.65$ & 2.81 & 20\\
 3 & 27.29 & 12.78 & $-14.51$ & 2.81 & 50\\
 4 & 49.56 & 38.71 & $-10.85$ & 2.22 & 30\\
 5 & 52.76 & 40.18 & $-12.58$ & 2.81 & 10\\ 
 6 & 55.88 & 46.27 & $-9.61$ & 2.50 & 20\\
 7 & 56.98 & 44.45 & $-12.53$ & 2.50 & 100\\
 8 & 57.99 & 45.42 & $-12.57$ & 2.50 & 10\\
 9 & 60.09 & 52.87 & $-7.22$ & 2.22 & 50\\
 10 & 62.21 & 50.46 & $-11.75$ & 1.98 & 200\\
\end{tabular}
\label{tab:VDMC-VOA}
\end{table}

\subsubsection{Destructive testing and damage mechanism}

After collecting all the data, we have performed a destructive test at the programmed setting of $63~\deci\bel$ (measured $46.9~\deci\bel$). Increasing power between 2.8~W and 4.4~W over $10~\second$ did not affect attenuation any further. Subsequently raising it to $4.5~\watt$~($36.5~\deci\bel\milli$) initiated fiber fuse at the flat connector~(FC)~interface between VDMC~VOA output and the beamsplitter arm. As a result, the beamsplitter-side fiber was burnt while VDMC~VOA output connector remained seemingly undamaged. After re-splicing the damaged section, cleaning the connector and repeating the experiment we observed the fuse again. Therefore this power setting may be considered a limit in realistic conditions. In order to study high exposure effects on the active element itself, we proceeded with the tests, this time directly splicing the beamsplitter arm to VDMC VOA output. Repeatedly applying to the same spot $4.5$--$6.8~\watt$ for $10~\second$ (with roughly 0.5~W step) resulted in dramatic drop in attenuation of $-20.7~\deci\bel$ in addition to the previous $-11.3~\deci\bel$, totaling $-32~\deci\bel$ decrease from the initial value. After this point 6.8~W power was repeatedly applied for 30, 60, 180, 300 and $600~\second$ without switching off the EDFA. Measurements made after the first two pulses indicated  further slight decremental changes ($-1.25~\deci\bel$ and $-1.07~\deci\bel$), the third and the fourth had virtually no effect ($<-0.1~\deci\bel$), while the last led to a slight attenuation increase ($0.48~\deci\bel$) probably due to structural damage of the glass disk surface. Neither of these tests resulted in breaking the active element and causing denial of service.

Mechanically disassembling VDMC VOA and analyzing the effects on the active element confirmed the expected damage mechanism of partial metal coating ablation during laser damage. As can be seen in~\cref{fig:VDMC-VOA}, concentric structures appear at the damage points after high power exposure. A closer look reveals that they consist of darker central parts and several brighter outer rings. Formation of such structures is typical for laser ablation processes in glasses with metal films or nanoparticles in the near-surface layer~\cite{domke2014,egorov2014,egorov2015}. They appear when glass and metal particles are evaporated and melted away from the center of affected area to cooler outer regions. We also observed cracks on the glass surface around some of the damaged regions (look like X-wing starfighters in~\cref{fig:VDMC-VOA}), caused by overheating and local surface defects. The proposed mechanism explains why attenuation decrease is less pronounced at the sample regions with lower attenuation, where the metal film is initially much thinner (see \cref{tab:VDMC-VOA}). The minimal achieved loss is limited by insertion loss of the glass disk and other attenuator components, whose experimentally measured value was 1.7~dB.

\section{Statistical risk estimate}
\label{sec:statistics}

It would be common to say at the conclusion of this study that other QKD systems may be vulnerable to the laser-damage attack as the results of this Article make clear. However, is this a practically significant risk or not? A more precise statistical statement can be made using methods that are not familiar to most readers in the physics field, but are much more common in other fields that study complex systems. In these fields, the likelihood of outcomes is often estimated without a detailed understanding of the underlying processes. Clinical trials of new drugs and prediction of the magnitude and frequency of natural disasters are examples. Here we make an observation that the laser damage attack on QKD is a similar setting, and attempt to make a relevant statistical prediction using the methodology borrowed from outside the quantum physics field.

The laser damage belongs to a class of attacks that are not based on a clear physical model. The outcome of such attack on a particular QKD system -- either denial of service or security breach -- depends on a multitude of factors difficult to predict in advance, and is only possible to ascertain by experimental testing. (Other examples of attacks in this class are detector control attacks~\cite{makarov2006,lydersen2010a,lydersen2010b,wiechers2011,weier2011,lydersen2011b,lydersen2011c,sauge2011,sajeed2015a,huang2016} and timing attacks~\cite{lamas-linares2007,nauerth2009,xu2010,jain2011,huang2018}. To give a contrary example, a photon number splitting attack~\cite{felix2001,brassard2000} has a well understood physical model, and its outcome can be predicted theoretically based on system design and specifications.)

In this setting, a risk prediction for untested QKD systems can be made by Bayesian analysis \cite{gelman2014} after testing a small subset of the systems. A crucial assumption here is that this subset is chosen among the entire system population at random. This assumption is currently not possible to enforce with industrial QKD systems, because many of them would not be made available for testing if requested by us. However, we feel that our choice of the samples for testing has been sufficiently wide ranging, and we have not excluded any system from the testing based on our expectations or its design. With this caveat, we apply the Bayesian analysis.

After two QKD systems were tested and compromised by the laser-damage attack in Ref.~\onlinecite{makarov2016}, a statistical prediction was made that a significant fraction ($>20\%$) of remaining untested QKD implementations were almost certain ($99.0\%$ probability) to contain similar unpatched loopholes. This was a practically relevant prediction: if one-fifth of the existing systems were almost certain to be vulnerable to the attack, then this was a serious security risk, and the community needed to worry about this attack.

The outcome of our present study is consistent with this prediction. We remark that we have only tested one component (optical attenuator) in one half of each QKD system (the source) against one type of laser illumination ($< 9~\watt$ c.w.\ at $1550~\nano\meter$). Eve could in principle try higher power, pulsed illumination regimes and different wavelengths to improve her attack, and she could try attacking the receiver side as well. Even with our restricted testing, we have observed at least 2 attenuator types and thus QKD systems that use them compromised.

Let's update the Bayesian analysis prediction, taking into account the additional outcomes from our present study. For the sake of this discussion, we give the QKD system containing the fixed attenuator a big benefit of doubt, and assume the outcome to be the denial of service. The systems containing the MEMS and VDMC VOAs are considered to be compromised. The system with the manual VOA is excluded from the statistics so far, because a higher laser power is needed to complete its testing. This leaves us (taking the two original data points \cite{makarov2016} into account) with 5 systems tested to the date, out of which 1 denial-of-service and 4 security compromise outcomes have been observed. The Bayesian analysis prediction then gives $99.5\%$ probability that $>20\%$ of the remaining untested systems are vulnerable \footnote{The probability is given by beta cumulative function (Eq.~12 in Ref.~\onlinecite{murphy2007} but integrated from $0$ to $t$, where $t$ is the fraction of untested samples that can be safe to the attack while having at least $1-t$ of untested samples vulnerable). We assume the entire population comprises 50 QKD systems. We assume Jeffreys prior, although with five outcomes the choice of prior affects the prediction already very little.}. This confirms that the laser-damage attack on today's QKD implementations should be taken very seriously.

\section{Countermeasures}
\label{sec:countermeasures}

Our testing has shown that none of the attenuators is confidently robust against the high-power laser damage attack (the test of manual VOA is inconclusive owing to the insufficient power of our EDFA). Thus, countermeasures to this attack need to be developed and tested. A straightforward countermeasure, a watchdog monitor, might not be sufficient, as the monitor itself can be destroyed under high power~\cite{makarov2016}. Instead, passive countermeasures like optical isolators and circulators are used in some QKD systems (including commercial ones) to add isolation between the vulnerable attenuator and the quantum channel~\cite{lucamarini2015,qctek2019}. While this may protect the vulnerable attenuator from the laser damage attack, the isolators and circulators also need to be tested for laser damage. This follow-up study is under way \cite{ponosova2020}.

Alternatively we propose to add a special passive component, an optical fuse, at the source's output. The optical fuse will only tolerate a certain amount of laser power, and will permanently disconnect itself once injected power crosses a threshold. In this way, this fuse physically interrupts the injected high power and protects the system from the laser damage attack. Such a device has been previously proposed using a TeO\textsubscript{2} soft glass segment inserted inline a standard fiber, which prevents pulses higher than approximately $1~\watt$ (with duration $\sim 1~\second$) from passing through~\cite{todoroki2004}. Adopting this or similar technique into a QKD system could be another future study.

\section{Conclusion}
\label{sec:conclusion}

In this work, four types of fiber-optic attenuators commonly used in QKD implementations have been tested under high-power continuous-wave laser. The manual variable attenuator exhibits minimal change during testing. The fixed attenuator exhibits a temporary drop in attenuation at $2.5~\watt$ ($34~\deci\bel\milli$) of optical power. The MEMS VOA and VDMC VOA both show a permanent and large -- several decibel -- decrease in attenuation. The decreased attenuation results in the increased intensity of transmitted states, which can be exploited by Eve to compromise the security of QKD. This shows that the mean photon number can be tampered with, which effectively breaks the fundamental assumption about the mean photon number crucial in a QKD system with a weak coherent source~\cite{gottesman2004,ma2005}. Our study also confirms earlier statistical predictions about the danger of the laser damage attack on QKD.

The demonstrated attack shows one more way to break the fundamental assumption about mean photon numbers in the QKD security proofs, in addition to a laser seeding attack~\cite{huang2019}. A detailed analysis of its effect on decoy-state BB84 and MDI QKD protocols is given in Ref.~\onlinecite{huang2019}. We hope our work encourages the development of non-leaky state preparation in MDI and device-independent QKD systems.

\acknowledgments

We thank our industry collaborators for their cooperation. We thank J.-P.~Bourgoin for recalculating the Bayesian analysis. This work was funded by NSERC of Canada (programs Discovery and CryptoWorks21), CFI, MRIS of Ontario, the National Natural Science Foundation of China (grants 61901483, 61601476, and 61632021), the National Key Research and Development Program of China (Grant 2019QY0702) and the Ministry of Education and Science of Russia (programs 5-in-100 and NTI center for quantum communications). A.H.\ was supported by China Scholarship Council. This work was funded by Government of Russian Federation (grant 08-08).

{\em Author contributions:} A.H.,\ R.L.,\ and V.E.\ conducted the experiments and analysed the data. S.T.\ designed and built the custom EDFA, and K.K.\ the fiber fuse monitor. A.H.,\ R.L.,\ V.E.,\ and V.M.\ wrote the Article. V.M.\ supervised the study.

\def\bibsection{\medskip\begin{center}\rule{0.5\columnwidth}{.8pt}\end{center}\medskip} 

%

\end{document}